\documentstyle[12pt]{article}
\newfont{\gothic}{eufm10 scaled\magstep 1}

\def\pd{\partial}
\begin{document}
\begin{flushright}
 DTP-98-43
\end{flushright}
\vskip 0.8 cm

\begin{Large}
\centerline{{\bf Moyal Brackets, Star Products}}
\vskip 0.1cm
\centerline{{\bf and the Generalised Wigner Function.}}
\end{Large}
\vskip 1.0cm

\centerline{\ \ David B. Fairlie$^\dag$}
\vskip 0.5cm

\centerline{{$^\dag$}Department of Mathematical Sciences,}
\vskip 5pt
\centerline{University of Durham,}
\vskip 5pt
\centerline{Durham, DH1 3LE, England.}
\vskip 5pt
\centerline{{\sl david.fairlie@durham.ac.uk}} 
\vskip 10pt
\begin{abstract}
The Wigner-Weyl- Moyal approach to Quantum Mechanics is recalled, and similarities to classical probability theory emphasised. The Wigner distribution function is generalised and viewed as a construction of a bosonic object, a target space co-ordinate, for example, in terms of a  bilinear convolution of
two fermionic objects, e.g. a quark antiquark pair. This construction is essentially non-local, generalising the idea of a local current. Such Wigner functions are shown to solve a BPS generalised  Moyal-Nahm equation.
\end{abstract} 
\section
{Introduction}
The intention of this article is to argue for the unique and universal appearance of the associative star product in the process of quantisation and to emphasise its essentially non local structure.
The construct upon which everything depends is the unique associative product on functions defined over a symplectic space, the star product. On a two dimensional phase space  $(x,p)$ it may be defined for functions $f(x,p),\ g(x,p)$ by

\begin{equation}
f\star g = \exp\kappa  \Bigg[ 
\frac{\partial~}{\partial x}\frac{\partial~}{\partial\tilde p}-
\frac{\partial~}{\partial p}\frac{\partial~}{\partial\tilde x}
\Bigg] f({\bf x}) g({\bf \tilde x}) \vert_{{\bf x} = {\bf\tilde x}}
\label{star1}
\end{equation}

\noindent with ${\bf x}=(x,\ p)$ so

\begin{equation}
f\star g = \sum_{s=0}^\infty \frac{\kappa^s}{s!} \sum_{t=0}^s
(-1)^t \pmatrix{ s \cr t \cr}
[\partial_x^{s-t} \partial_p^t f]
[\partial_x^t \partial_p^{s-t} g]\,.
\label{star2}
\end{equation}

This definition of the star operator may be extended to a $2 N$ dimensional phase space with canonically conjugate variables $(x_j,\ p_j)$  as follows;

\begin{equation}
f\star g\ = f{\rm e}^{\frac{i\hbar }{2}\sum_{j=1}^N(\stackrel{\leftarrow }{\partial
}_{x_j} \stackrel{\rightarrow }{\partial }_{p_j}-\stackrel{\leftarrow
}{\partial }_{p_j}\stackrel{\rightarrow }{\partial }_{x_j})} g
\label{perhaps} 
 \end{equation}
or,equivalently
\begin{equation}
f\star g = \sum_{j=0}^N\sum_{s=0}^\infty \frac{\kappa^s}{s!} \sum_{t=0}^s
(-1)^t \pmatrix{ s \cr t \cr}
[\partial_{x_j}^{s-t} \partial_{p_j}^t f]
[\partial_{x_j}^t \partial_{p_j}^{s-t} g]\,.
\label{star3}
\end{equation}
The crucial feature of the star product, besides its associativity and uniqueness \cite{bayen}\cite{fletch}, is the property
\begin{equation}
\int\int\dots\int^\infty_{-\infty}f(x_i,p_i)\star g(x_i,p_i) \prod dx_j \prod dp_j =
\int\int\dots\int^\infty_{-\infty}f(x_i,p) g(x_i,p_i)\prod dx_j \prod dp_j.
\label{unique}
\end{equation}
 This property follows directly from another representation of the star product
\cite{baker}
\begin{equation}
f\star g =\frac{1}{\kappa^{2N}}\int \sum_i{\rm e}^{\frac{i}{\kappa}\det\left|\begin{array}
{ccc}1&1&1\\
     x_i&x'_i&x''_i\\
     p_i&p'_i&p''_i\end{array}\right|}f(x'_j,p'_j)g(x''_j,p''_j)\prod dx'_k\prod dx''_k\prod dp'_k\prod dp''_k
\label{baker}
\end{equation}
The validity of the representation itself can be seen from performing a Fourier resolution of $f,\ g$. What it means is that all terms in the star product except the first, $fg$, are divergences and so integrate out provided $f,\ g$ and their derivatives vanish sufficiently strongly at infinity.
The Moyal bracket which is proportional to the imaginary part of the star product with parameter $\kappa= \frac{\hbar}{2}$, with proportionality constant
$\frac{2}{\hbar}$ first made its appearance in Quantum Mechanics, through Moyal's \cite{moyal} equation for the Wigner function, 
\begin{equation}
\frac{\partial~}{\partial t}f(x,p,t) = H\star f-f\star H\label{moyal}
\end{equation}
\section{ `Physical' motivation for Moyal quantisation.}

Suppose $H(x,p)$  is  distributed with normalised probability distribution $f(x,p)$
i.e the probability of finding $H(x,p)$ in the range $x~ {\rm to}~  x+dx,\ p~   {\rm to}~ p+dp$ is  $H(x,p)f(x,p)dxdp$
Then classically, the expectation of the energy, i.e. the average energy is given by
\begin{equation}
E=\frac{\int_\infty^\infty\int_\infty^\infty H(x,p)f(x,p)dxdp}{\int_\infty^\infty\int_\infty^\infty f(x,p)dxdp}
\label{prob}
\end{equation}
where the denominator is a normalisation factor to guarantee that $\displaystyle{\frac{f(x,p)}{\int_\infty^\infty\int_\infty^\infty f(x,p)dxdp}
}$
will behave as a probability.
What is the quantum mechanical equivalent of this statement? Notice that the equation (\ref{prob}) can be written in the equivalent form  
\begin{equation}
\int_\infty^\infty\int_\infty^\infty H(x,p)*f(x,p)dxdp
=E\int_\infty^\infty\int_\infty^\infty f(x,p)dxdp
\label{starprob}
\end{equation}
provided that $f(x,p)$ vanishes at infinity. Now in the mathematization of physics, the physical law is frequentely presented in its most intuitive aspect as  an integral law, which is  then converted into the more mathematically manageable form of a differential law by means of an integral theorem. One outstanding example is Gauss' Law, which says that the integral of the normal component of the electric field  over a surface is proportional to the total electric charge within the surface;
\begin{equation}
\int_S{\bf E.dS} =4\pi \int_V \rho dV.
\label{gauss}
\end{equation}
Use of Gauss law enables the left hand side to be converted to a volume integral
$\int_v \nabla.{\bf E}dV$. Since $V$ is arbitrary we can equate integrands to 
obtain Poisson's equation,
\begin{equation}
{\rm Div}{\bf E} =4\pi \rho.
\label{pois}
\end{equation}
 which is a differential relationship.
Another example is the standard inference of the existence of the representaion of a force field ${\bf F}$ as the negative gradient of a potential which follows from the observation that the work done against the force in any closed circuit is zero; i.e. $\oint {\bf F.dr}=0$, an integral law.

While it is not a necessary consequence of the integral law (\ref{starprob})
that we can equate integrands on both sides, as contrary to the previous 
examples, the integration region cannot be chosen arbitrarily, let us pursue the consequences of so doing. After all, we must intoduce some illogical postulate if we hope to  infer a quantum result from a classical one. This procedure has some echoes of looking for local gauge transformations.
We shall demonstrate a result, already in the literature, but not apparently widely appreciated, that the equations which result from equating integrands
\begin{equation}
H(x,p)*f(x,p)=E f(x,p)
\label{quant}
\end{equation}
are precisely equivalent to ordinary, time independent quantum mechanics.

In a slight generalisation of Wigner's original proposal \cite{wigner},
 diagonal ($a=b$) and non-diagonal ($a\neq b$) Wigner functions $f_{a,b}$ are defined
by 
\begin{equation}
f_{ab}(x,p)={\frac{1}{2\pi }}\int \!dy~\psi _{a}^{*}(x-{\frac{\hbar }{2}}
y)~e^{-iyp}\psi _{b}(x+{\frac{\hbar }{2}}y)\ .\label{wig}
\end{equation}
They are ``self-orthogonal'' upon taking a phase-space ``trace'' 
\begin{equation}
\int dx\,dp\,f_{ab}(x,p)=\,\delta _{ab}\;,\label{ort}
\end{equation}
assuming the wave functions are orthonormal, $\int dx\psi _{a}^{*}(x)\psi
_{b}(x)=\delta _{ab}$. When the wave functions are energy eigenfunctions,
the $f$'s satisfy the two-sided energy $\star $-genvalue equations in the terminology of \cite{cos}
\begin{equation}
H\star f_{ab}=f_{ab}\,E_{b}\;,\;\;\;f_{ab}\star  H=E_{a}\,f_{ab}\;,\label{stargen}
\end{equation}
These are just the  equations (\ref{quant}).   

The proofs of these results follow easily from the following
Lemma\footnote{This was suggested by an unpublished result of Ian Strachan}\rlap. 
\begin{eqnarray}
{\rm e}^{py}f(x)\star{\rm e}^{py'}g(x)&=&
\sum_j\sum_k{\rm e}^{p(y+y')}(-\lambda)^{m+n}\frac{(y'\partial_x)^m}{m!}f(x)
\frac{(-y\partial_x)^n}{n!}g(x)\nonumber\\
&=&{\rm e}^{p(y+y')}f(x+y')g(x-y)
\label{id}
\end{eqnarray}
The time dependent equations \cite{dbf2}  can be expressed in the same form as  
the time inedendent ones (\ref{quant}) provided $H$ is replaced by $H'=H-Et$
and the star product is taken with the pairs $x,\ t;\ p,\ E$ according to
(\ref{star3}). This avoids the need to introduce a second evolutionary parameter, as was done in \cite{dbf2}.

\section{Moyal Brackets in String Physics}
There is a second way in which Moyal Brackets make their appearance in physics;
since they are closer to a commutator than the Poisson Brackets, they can effect the transition between a matrix valued field theory and  the $N\rightarrow\infty$ limit of that theory \cite{cosmas1}\cite{cosmas2}.
In recent years the study of super Yang Mills dependent upon a single evolutionary parameter for a gauge group $SU(N)$ has become fashionable as a paradigm for the extraction of information about M-theory; \cite{martyr}\cite{banks}\cite{dvv}\cite{fa1}\cite{fai}\cite{castro}\cite{douglas}. In one example, with 8 transverse directions, in a gauge with $X^8$ set to zero, the bosonic sector of this theory admits a set of first order Bogomol'nyi equations, which are just an expression of self duality in 8-dimensions \cite{cor};
\begin{eqnarray}
\frac{\pd ~}{\pd\tau}X^1&=& [X^2,\ X^7]+[X^6 ,\ X^3]+[X^5,\ X^4]\nonumber\\
\frac{\pd ~}{\pd\tau}X^2&=&[X^7,\ X^1]+[X^5 ,\ X^3]+[X^4,\ X^6]\nonumber\\
\frac{\pd ~}{\pd\tau}X^3&=&[X^1,\ X^6]+[X^2,\ X^5] +[X^4,\ X^7]\nonumber\\
\frac{\pd ~}{\pd\tau}X^4&=&[X^1,\ X^5]+[X^6 ,\ X^2]+[X^7,\ X^3]\label{nahm10}\\
\frac{\pd ~}{\pd\tau}X^5&=&[X^4,\ X^1]+[X^3 ,\ X^2]+[X^6,\ X^7]\nonumber\\
\frac{\pd ~}{\pd\tau}X^6&=&[X^3,\ X^1]+[X^2 ,\ X^4]+[X^7,\ X^5]\nonumber\\
\frac{\pd ~}{\pd\tau}X^7&=&[X^1,\ X^2]+[X^3 ,\ X^4]+[X^5,\ X^6]\nonumber.
\end{eqnarray}

These are just the Nahm equations in 7-dimensions. The corresponding second order equations are just the Yang Mills Equations, thanks to the Bogomol'nyi
property. The continuum limit of these equations in which the matrices $X^\mu$ are replaced by functions $X^\mu(x,p,\tau)$ are obtained by replacing the commutators by Poisson Brackets. This limiting procedure may be approached
by replacing the commutators, not by a Poisson Bracket, but rather a Moyal
Bracket. It was first pointed out in \cite{cosmas2} that the Poisson Bracket continuum version of the squared commutator in the Yang Mills action is just the Schild version of the action for a classical string. It turns out that in the Moyal bracket formalism the equations may be solved in terms of a generalisation of the Wigner function. The generalisation proposed is expressed in a notation suggestive of Dirac terminology, but this is only analogical, as the `spinors'
below depend only essentially upon one variable, and are better thought of as
simply column vectors.
 
\subsection{Multicomponent Generalisation}
Suppose $\gamma^{j},\ j=1\dots 7$ are 7 gamma matrices   which admit a  representation by 7 real antisymmetric $8\times2^{8}$ matrices as follows
\begin{eqnarray}
\gamma^1&=&i\sigma_1\otimes\sigma_3\otimes\sigma_2\nonumber\\
\gamma^2&=&i\sigma_3\otimes\sigma_2\otimes\sigma_3\nonumber\\
\gamma^3&=&i\sigma_1\otimes\sigma_2\otimes I\!\! I\nonumber\\
\gamma^4&=&i\sigma_3\otimes I\!\! I\otimes \sigma_2\label{gmats}\\
\gamma^5&=&i\sigma_2\otimes I\!\! I\otimes I\!\! I\nonumber\\
\gamma^6&=&i\sigma_3\otimes\sigma_2\otimes\sigma_1\nonumber\\
\gamma^7&=&i\sigma_1\otimes\sigma_i\otimes\sigma_2\nonumber
\end{eqnarray}
and $\psi$ is an $8$ component spinor,  Then one can construct a generalisation of the Wigner distribution
 \begin{equation}
 X^k=i\int_{-\infty} ^\infty {\psi^\dag}(x-y,\tau)\gamma^k\psi(x+y,\tau){\rm e}^{\frac{2i\pi py}{\lambda}}dy,\ k=1\dots 7
\label{gen}
\end{equation}
\begin{eqnarray}
X^j\star X^k&=&\int_{-\infty}^\infty\int_{-\infty}^\infty
{\psi^\dag}(x-y,\tau)\gamma^j\psi(x+y,\tau)
{\rm e}^{\frac{2i\pi py}{\lambda}}\star{\psi^\dag}(x-y',t)\gamma^k\psi(x+y',\tau)
{\rm e}^{\frac{2i\pi py'}{\lambda}}dydy'\nonumber\\
&=&\int_{-\infty}^\infty\int_{-\infty}^\infty
{\psi^\dag}(x-y+y'),\tau)\gamma^j{\psi^\dag}(x+y+y',\tau)
{\rm e}^{\frac{2i\pi py}{\lambda}}{\psi^\dag}(x-y'-y,t)\gamma^k\psi(x+y'-y,\tau)
{\rm e}^{\frac{2i\pi py'}{\lambda}}dydy'\nonumber\\
&=&Z(\tau)\int_{-\infty}^\infty{\psi^\dag}(x-y-y',\tau)\gamma^k\gamma^j\psi(x+y+y',\tau){\rm e}^{\frac{2i\pi p(y+y')}{\lambda}}d(y+y')\label{zeqn}\end{eqnarray}
where an integration over the variable $y-y'$ has been performed and orthogonality of the spinors ${\psi^\dag}(x,\tau),\ \psi(x,\tau)$ has been assumed in the form 
\begin{equation}
\int^\infty_{-\infty}{\psi^\dag}(x,\tau)_\alpha,\ \psi(x,\tau)_\beta dx=
Z(\tau)\delta_{\alpha\beta}.\label{ortho}
\end{equation}
If additionally the structure of the components of $\psi(x,\tau)$ takes the form
\begin{equation}
\psi(x,\tau)_\mu =\phi(x)_\mu f(\tau),\ \ \mu=1,\dots,7,\ \ \psi(x,\tau)_8 =\phi(x)_8 f^*(\tau),\ \ Z(\tau)=|f(\tau)|^2
\end{equation}
Then $\displaystyle{\frac{\pd ~}{\pd\tau}X^\mu}$ will take the form
\begin{eqnarray}
\frac{\pd ~}{\pd\tau}X^\mu&=&\frac{\pd ~}{\pd\tau}|f(\tau)|^2\int_{-\infty} ^\infty {\phi^\dag}(x-y)\gamma^\mu\phi(x+y){\rm e}^{\frac{2i\pi py}{\lambda}}dy\nonumber\\
&+&\frac{\pd ~}{\pd\tau}(f^2-{f^*})^2\sum_{j=1}^{j=7}\int_{-\infty} ^\infty {\phi^\dag}(x-y)_j\gamma^k\phi(x+y)_8{\rm e}^{\frac{2i\pi py}{\lambda}}dy,\label{derive}
\end{eqnarray}
and it is possible to show that our ansatz will solve the Moyal-Nahm equations. These equations are just those of (\ref{nahm10}) with commutators replaced by Moyal brackets.
The antisymmetry of the representations of the 7 gamma matrices has been assumed. This ensures that the $X^\mu$ are real. This choice of $\tau$ dependence is motivated by further developments.

Now suppose that  $\displaystyle{Z(k)=|f(\tau)|^2 = \frac{\pd ~}{\pd\tau}|f(\tau)|^2=Z(k)^2=|f(\tau)|^4}$
This will be the case if $f(\tau)$ has the form $\displaystyle{f(\tau)=\frac{{\rm e}^{i\theta}}{\sqrt\tau}}$, where $\theta$ is an  arbitrary constant. Suppose we now wish to
solve the Moyal form of (\ref{nahm10})
\begin{equation}
\frac{\pd ~}{\pd\tau}X^1= \{X^2,\ X^7\}_{MB}+\{X^6 ,\ X^3\}_{MB}+
\{X^5,\ X^4\}_{MB},
\label{example}
\end{equation}
et cetera. 
The right hand side of this sample equation is the matrix element (in the sense of the Wigner construction) of twice the sum $\gamma_2\gamma_7+\gamma_6\gamma_3+\gamma_5\gamma_4$,
as the $\gamma$'s are anti-symmetric. Now
\begin{equation}
\gamma_2\gamma_7+\gamma_6\gamma_3+\gamma_5\gamma_4 = \gamma_1+ 4(\delta_{8,2}-\delta_{2,8})\label{find}
\end{equation}
In other words the matrices on the left and right of the typical equation differ only in their last rows and columns. This is true for all seven equations. This fact may be exploited in an expeditious choice of the constant $\theta$ to assert that the extended Wigner functions (\ref{wig}) with the choice of
Dirac matrices for $\gamma^k$  make (\ref{example}) to be identically satisfied by arranging that 
\begin{eqnarray}
&~&\frac{\pd ~}{\pd\tau}i\int_{-\infty} ^\infty {\psi^\dag}(x-y,\tau)\gamma^k\psi(x+y,\tau){\rm e}^{\frac{2i\pi py}{\lambda}}dy
\nonumber\\
&=&iZ(\tau)\int_{-\infty} ^\infty {\psi^\dag}(x-y,\tau)[\gamma^1+4(\delta_{8,2}-\delta_{2,8})]\psi(x+y,\tau){\rm e}^{\frac{2i\pi py}{\lambda}}dy\label{require}
\end{eqnarray}
with similar equations for the other components $X^\mu$. This will be arranged if $\theta$ is chosen to satisfy
\[\sin(2\theta)=\frac{\sqrt3}{2}\]
i.e. $\theta = \frac{\pi}{6}$
Thus it is possible to solve the Moyal Nahm equations, albeit with a somewhat trivial time dependence in terms of a generalised Wigner distribution. By construction, the solution almost, but not quite retains the full Lorentz invariance of the original equations. It fails only in the fact that the last component of $\psi$ has been treated differently from the others. There is in fact no compelling reason that in the generalisation of the Wigner function proposed here that the matrices involved should be anticommuting gamma matrices,
apart from the retention of as much Lorentz invariance as possible.
Other choices may open up new possibilities for bilinear equations which admit a solution in terms of Wigner functions. For example, as a spin-off from these considerations, integrable `Euler' top equations in dimensions $2^N-1$ have been discovered \cite{ueno}\cite{ueno2}. 
\section{Discussion}
The generalised Wigner functions, like the original are real valued, hence their interpretation as target space co-ordinates is viable. As a bilinear construction  in $\psi$ and $\psi^\dag$ it may be viewed as a non-local extension of a current $\bar\psi\gamma^\mu\psi$, and the physical interpretation as a quark-antiquark pair bound together is appealing. The fact that these generalised functions solve the Bogomol'nyi equations associated with
the Matrix String equations can be viewed as an extension of the bilinear parametrisation of the string equations given in \cite{manog}.

\end{document}